\documentclass{PoS}
\usepackage{graphicx}
\PoS{PoS(LAT2005)121}
\newcommand{\srm}[1]{{\textrm{\scriptsize{#1}}}}
\newcommand{\DCI}{\ensuremath{D_\srm{CI}\ }}

\title{Low energy constants from the chirally improved Dirac operator \DCI 
\footnote{For the Bern-Graz-Regensburg (BGR) Collaboration.}}

\ShortTitle{Low energy constants from the chirally improved Dirac operator \DCI}

\author{Christof Gattringer\\
Karl-Franzens-Universit\"at Graz, Austria\\
E-mail: \email{christof.gattringer@uni-graz.at}
}

\author{\speaker{Philipp Huber}\thanks{Supported by Fonds zur F\"orderung
der wissenschaftlichen Forschung in \"Osterreich, project P16823-N08.}\\
Karl-Franzens-Universit\"at Graz, Austria\\
E-mail: \email{philipp.huber@stud.uni-graz.at}
}

\author{C. B. Lang\\
Karl-Franzens-Universit\"at Graz, Austria\\
E-mail: \email{christian.lang@uni-graz.at}
}

\abstract{The leading order low energy parameters like the pion decay constant
or the  quark condensate are well-known from ``classical'' low energy theorems
and experiments. It is a challenge, however, to find these parameters based 
exclusively on an ab-initio QCD calculation. We discuss results of a quenched
lattice calculation of low energy constants using the chirally improved Dirac
operator.  Several lattice sizes at different lattice spacings are studied,
using pseudoscalar and axial vector correlators. We find consistent results for 
$f_\pi = 96(2)\;\textrm{MeV}$, $f_K = 105(2)\;\textrm{MeV}$,  $\Sigma=
-(286(4)\; \textrm{MeV})^3$, the average light quark mass  $\bar m =
4.1(2.4)\;\textrm{MeV}$ and  $m_s = 101(8)\;\textrm{MeV}$. }

\FullConference{XXIIIrd International Symposium on Lattice Field Theory\\

                 25-30 July 2005\\

                 Trinity College, Dublin, Ireland}

\begin{document}

\section{Setup}
When QCD was accepted as the quantum field theory of strong intereractions it
became clear that its (approximate) chiral flavor symmetry is spontaneously
broken. The exploitation of the underlying principles has then led to the
development of chiral perturbation theory. In that context a systematic
expansion of many observables in terms of low energy constants has been derived.
These constants, however, have to be determined either from experiment or from
basic principles, i.e., non-perturbative solution of the underlying field theory
QCD. The lattice formulation of QCD allows such a determination. 

Since chiral symmetry and its spontaneous and explicit breaking play an
essential role, it is desirable to work with fermionic actions having that
symmetry. Lattice Dirac operators obeying the Ginsparg-Wilson relation (GWR) are
the lattice analogue of chirally symmetric continuum Dirac operators. Here we
use the chirally improved Dirac operator \ensuremath{D_\textrm{\scriptsize
CI}}\  \cite{Ga01GaHiLa00}. It is based on a systematic expansion of the lattice
Dirac operator taking into account the whole Clifford algebra and terms coupling
fermions within a certain range of neighbors on the lattice. The expanded Dirac
operator is inserted in the GWR which  then leads to a set of algebraic
equations for the expansion coefficients. In recent applications
\cite{GaGoHa03a}, as well as here, we use a set of 19 independent terms in the
action. Finally, in the  definition of \ensuremath{D_\textrm{\scriptsize CI}}\ 
also one step of HYP-smearing of the gauge configuration is included. For the
gauge fields we use the L\"{u}scher-Weisz  action. The lattice spacings have
been determined  using the Sommer parameter. We summarize the simulation
parameters in Table~\ref{tab:simulationparameters}.

\begin{table}[b]
\begin{center}
\begin{tabular}{rcccccr}
$ L^3\times T $   & $\beta$ & $a[\textrm{fm}]$ & $a[\textrm{GeV}^{-1}]$ & \#\,cf.\ & Type & $a\,m$ $(a\,m_s)$ \\ \hline
$8^3\times24$     & $7.90$  & $0.148$ & $0.750$       & 200 	 & $p,n$    & $0.02-0.20$	\\
$12^3\times24$    & $7.90$  & $0.148$ & $0.750$       & 100 	 & $p,n$    & $0.02-0.20$	\\
$12^3\times24$    & $8.35$  & $0.102$ & $0.517$       & 100 	 & $p,n$    & $0.02-0.20$	\\
$16^3\times32$    & $7.90$  & $0.148$ & $0.750$       &  99 	 & $p,n$    & $0.02-0.20$	\\
$16^3\times32$    & $8.35$  & $0.102$ & $0.517$       & 100 	 & $p,n$    & $0.02-0.20$	\\
$16^3\times32$    & $8.70$  & $0.078$ & $0.395$       & 100 	 & $p,n$    & $0.02-0.20$	\\ 
$16^3\times32$    & $7.90$  & $0.148$ & $0.750$       & 100 	 & $p,n,w$  & $0.02-0.20$ ($0.08, 0.10$)\\
$20^3\times32$    & $8.15$  & $0.119$ & $0.605$       & 100 	 & $p,n,w$  & $0.017-0.16$  ($0.06$)
\end{tabular}
\end{center}
\caption{Parameters of the simulation. Where the strange quark mass is  given
(in parenthesis), we also determined propagators for strange hadrons. Type
denotes the type of the quark source/sink and \#\,cf.\ the number of
configurations entering the analysis. \label{tab:simulationparameters}}
\end{table}

In Ref.\ \cite{GaGoHa03a} hadron masses for a quenched simulation based on
\ensuremath{D_\textrm{\scriptsize CI}} have been presented. Preliminary results
for low energy constants in that context have been published in Ref.\
\cite{GaGoHa03b}. At that time the renormalization constants relating these to
the continuum $\overline{\textrm{MS}}$-scheme were not available. Meanwhile the
necessary constants for quark bilinears  have been determined for
\ensuremath{D_\textrm{\scriptsize CI}}\ in \cite{GaGoHu04}. This now allows us
to compute some of the basic low energy parameters in the quenched case. For
this  purpose we study quenched QCD at various values of the quark masses and
determine results for $m_u=m_d\neq m_s$ in the $u \overline{d} $ and $u
\overline{s} $ meson sector for several lattice sizes and lattice spacings, down
to a pion mass of 330 \textrm{MeV}. 

All our results have been derived from correlation functions of pseudoscalar
interpolating fields $P= \overline{d} \,\gamma_5 \,u$ and $A_4 = \overline{d}
\,\gamma_4 \,\gamma_5 \,u$. Where we give physical, renormalized
($\overline{\textrm{MS}}$-scheme) values, we always use the renormalization
factors as obtained in the chiral limit, even in plots for non-vanishing bare
quark masses. The renormalization factors $Z$ relate to the continuum
$\overline{\textrm{MS}}$-scheme at a scale of $\mu= 2\;$GeV and are
$a$-dependent as given in \cite{GaGoHu04,GaHuLa05a}. We use the interpolating
fields $P$ and $A_4$ for both the pion (with degenerate light quark mass
$\overline m$), and the kaon, with one light and one strange quark. The bare
mass of the heavy quark is fixed with the help of the physical kaon mass. In
order to improve signals we use Jacobi-smeared sources and sinks for the quarks.

\section{Results}

In  Ref.\ \cite{GaHuLa05a} we discuss the normalization of the sources, methods
for fitting the propagators, error estimates (based on a jack-knife analysis),
numerical derivatives, and the results for the meson masses. Here we present
only some results for quark masses, condensate and the pseudoscalar decay
constants.

\subsection{Quark masses}

\begin{figure}[t]
\begin{center}
\includegraphics[width=8cm,clip]{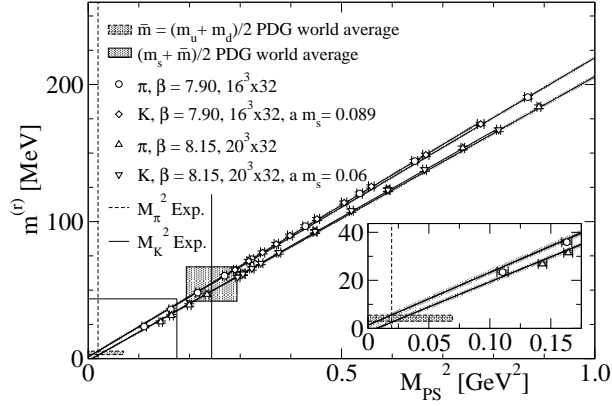}
\end{center}
\caption{$m^{(r)}$ vs.\ $M_{PS}^2$ using $X = P$. Here $M_{PS}$ denotes $M_\pi$ or
$M_K$, respectively. All masses are given in physical units in the $\overline{\textrm{MS}}$-scheme.}
\label{fig:mr_vs_MPS2}
\end{figure}

The renormalized quark mass may be determined, utilizing the axial Ward identity,
\begin{equation}
\label{eq:ratDAXPX}
\frac{Z_A}{Z_P}\,\frac{\langle \,\partial_t A_4(\vec p=\vec 0,t)\, X(0)\,\rangle}
{\langle\, P(\vec p=\vec 0,t)\,X(0) \,\rangle}
 \sim \frac{M_\pi^2\,f_\pi}{G_\pi^{(r)}}
= Z_m\,2\,m=2\,m^{(r)}
\;,
\end{equation} 
where $X$ may be $P$ or $A_4$. The asymptotic behavior of the correlators
cancels in this ratio and the plateau values provide the mass (proportional to
the so-called AWI-mass). For the second choice, $X=A_4$, the correlators are of
$\sinh$-type and the ratio becomes numerically unstable near the symmetry point
in $t$. Our analysis of the quark masses is based on the more stable choice
$X=P$.

The quark mass data in Fig.\ \ref{fig:mr_vs_MPS2}, presented in physical units
in the $\overline{\textrm{MS}}$-scheme, leads to a very consistent picture. The
abscissa gives the corresponding pseudoscalar mass, i.e., that of the pion or
the kaon, for the corresponding values of $m^{(r)}$. We find that for given
lattice spacing the numbers for kaon and pion are on top of each other, although
these states have different quark content.

The lines in Fig.\ \ref{fig:mr_vs_MPS2} correspond to linear fits 
$m^{(r)}\propto M_\pi^2$, enforcing the simultaneous chiral limit of both
observables. The data do not show deviation from that behavior, although
logarithmic corrections are expected due to quenching. The linear extrapolations
are in good agreement with the Particle Data Group \cite{PDBook04} average for
the light quark and strange masses at the physical pion mass. We obtain
\begin{equation}\label{eq:mlight}
\frac{1}{2}\left(m_u^{(r)}+m_d^{(r)}\right)\equiv\bar m^{(r)}
\simeq 4.1(2.4)\;\textrm{MeV}\,,\quad
\frac{1}{2}\left(m_s^{(r)}+\bar m^{(r)}\right)\simeq 52(3)\;\textrm{MeV}
\end{equation}
in the $\overline{\textrm{MS}}$-scheme. Combing theses values gives
$m_s^{(r)}=101(8)~\textrm{MeV}$. Possible finite size effects and other
systematic effects  like chiral extrapolation and quenching have not been
accounted for. The given error takes into  account the standard error and
includes the derivations due to the  dependence on the lattice spacing. The
numbers are in good agreement with determinations from the overlap action in 
\cite{GiHoRe02,ChHs03}.

\begin{figure}[t]
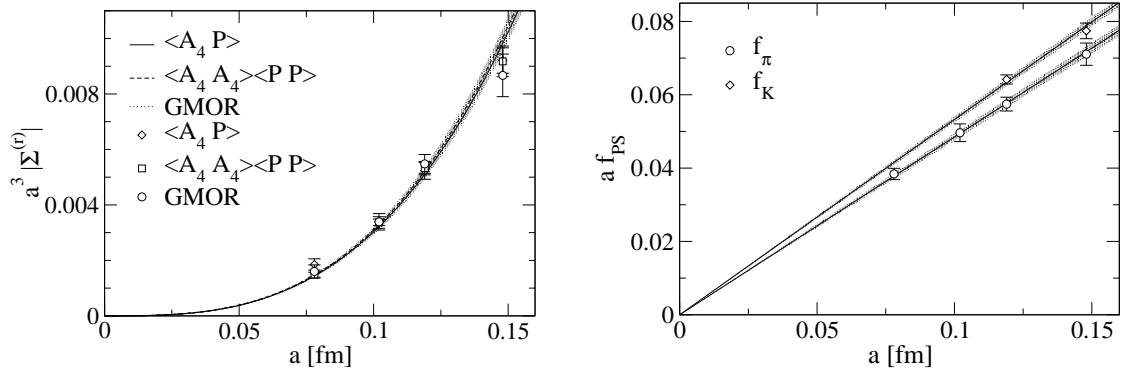

\begin{center}
\includegraphics[width=7cm,clip]{sigma_vs_a.eps}\qquad
\includegraphics[width=7cm,clip]{fPS_vs_a.eps}
\end{center}
\caption{\label{fig:scaling}Scaling properties of the light quark condensate
 (l.h.s.) and of the pion decay constant (r.h.s).}
\end{figure}

\subsection{Chiral condensate}

We have computed the renormalized condensate from the relations
\begin{eqnarray}
f_\pi^2\,M_\pi^2
& = & -2 \,m^{(r)} \,\Sigma^{(r)}\;,\\
Z_A\,Z_P\,\langle\, A_4(\vec p=\vec 0,t)\,P(0)\rangle 
\sim 
G_\pi^{(r)}\,f_\pi\, \mathrm{e}^{-M_\pi\,t}
=
\frac{f_\pi^2\,M_\pi^2}{2\,m^{(r)}}\, \mathrm{e}^{-M_\pi\,t}
&=& |\Sigma^{(r)}|\, \mathrm{e}^{-M_\pi\,t}\;,\\
Z_A\,Z_P\,\sqrt{ \langle\,A_4(\vec p=\vec 0,t)\, A_4(0)\rangle \,\langle\, P(\vec p=\vec 0,t)\,  P(0)\rangle}
&\sim&
|\Sigma^{(r)}|\,\mathrm{e}^{-M_\pi\,t}\;,
\end{eqnarray}
which all contain $\Sigma^{(r)}$. The first of these is the GMOR relation, the
other two are determinations directly from the coefficients  of propagators and
implicitly related to GMOR as well.

We find excellent agreement for all three determinations; the values are
consistent within the error bars.  The dependence on the bare quark mass is
compatible with the leading (linear) chiral behavior.  Note, that we are not at
small enough quark masses to be in the so-called $\epsilon$-regime
\cite{HeJaLe99} but are in the $p$-regime.

In Fig.\ \ref{fig:scaling} (l.h.s) we show the results of the linear
extrapolation to the chiral limit for all three types of determination for all
lattice sizes studied. The expected scaling behavior $\mathcal{O}(a^3)$ is
demonstrated by the fitted curves. Assuming the leading scaling behavior
throughout (as was done in other studies where only one lattice spacing was
studied) the average of the resulting values  for the condensate in the
continuum limit is $|\Sigma^{(r)}|=(286(4)~\textrm{MeV})^3$. This is slightly
larger than a determination from the overlap action \cite{WeWi05,GiHoRe02}
(although still with in the error limits) and larger than the corresponding
results in Ref.\ \cite{ChHs03,BiSh05}. Our numbers are in good agreement with
calculations in the $\epsilon$-regime \cite{GiHeLa04,FuHaOk05}.

\begin{figure}[t]
\begin{center}
\includegraphics[width=8cm]{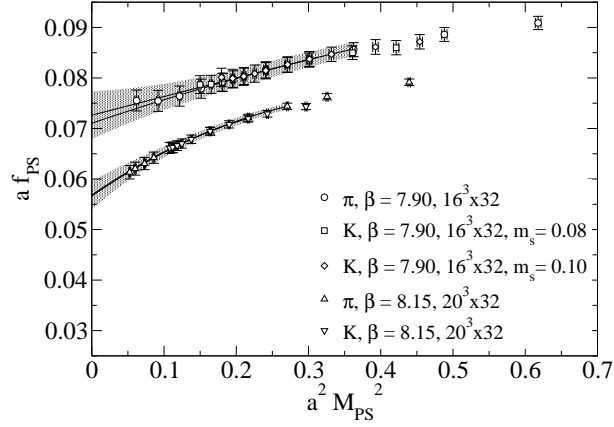}
\end{center}
\caption{\label{fig:afpiK_vs_a2mPS2} $a\,f_{\pi,K}$ vs.\ $(a\,M_{PS})^2$ with
error bands for chiral extrapolations as discussed in the text. The fit includes
only data in the indicated range.}
\end{figure}

\subsection{Decay Constants}

The pseudoscalar decay constants have been extracted from the asymptotic
behavior of the pseudoscalar correlation functions according to
\begin{equation}
\label{eq:corrAA}
Z_A^2\,\langle\, A_4(\vec p=\vec 0,t)\,A_4(0)\rangle 
\sim
M_{PS}\,f_{PS}^{\,2}\,\mathrm{e}^{-M_{PS}\,t}
\end{equation}
for pion and kaon, respectively.  When plotting them as functions of the
respective pseudoscalar masses in Fig.\ \ref{fig:afpiK_vs_a2mPS2}, the data for
pion and kaon essentially overlap each other and exhibit a universal functional
behavior. We also show the error band of a quadratic extrapolation to the chiral
limit. In quenched QCD one expects correction terms with a logarithmic
singularity in the valence quark mass $m$. As pointed out in \cite{Sh92}, the
leading order logarithmic term $m\, \log m$ of ChPT involves quark loops that
are absent in the quenched case. There will be non-leading, e.g., logarithmic,
terms, though.  In addition to the  term  linear in the quark mass $m$ we
therefore also allow a term $m^2 \,\log m$ in the extrapolating fit (cf.\ the
discussion in \cite{DoDrHo04}). Actually, in the fit it makes no significant
difference whether we take this term or just $m^2$.

In Fig.\ \ref{fig:scaling} (r.h.s.) we show the scaling behavior by comparing
results for different lattice constants for our largest physical size lattices.
The results are from the chiral extrapolation for the light quarks; in the
(semi-) chiral extrapolation for the kaon decay constant  the strange quark mass
parameter is held fixed. Assuming the leading $\mathcal{O}(a)$ behavior, the
continuum limit values are 
\begin{equation}
f_\pi=96(2)\;\textrm{MeV}\;,\quad
f_K=105(2)\;\textrm{MeV}\;.
\end{equation}
Studies for the overlap action in quenched simulations have obtained similar 
results for $f_\pi$ \cite{DoDrHo04,GiHeLa04,FuHaOk05,BiSh05,DuHo05,DaHeSp05}.
The experimental values are  $f_\pi= 92.4(0.3)\;\textrm{MeV}$ and 
$f_K=113.0(1.3)\;\textrm{MeV}$  \cite{PDBook04}.


\end{document}